\documentclass[aps,prl,onecolumn,superscriptaddress,bibnotes,preprintnumbers,amsmath,amssymb]{revtex4}
\usepackage{graphicx}% Include figure files
\usepackage{dcolumn}% Align table columns on decimal point
\usepackage{bm}% bold math
\usepackage{epstopdf}
\usepackage{mathtools}
\usepackage{natbib}
\usepackage{captcont}

\begin{document}
\title{A possible macronova in the late afterglow of the `long-short' burst GRB 060614}

\affiliation{Key Laboratory of Dark Matter and Space Astronomy, Purple Mountain Observatory, Chinese Academy of Sciences, Nanjing 210008, China }
\affiliation{University of Chinese Academy of Sciences, Yuquan Road 19, Beijing, 100049, China}
\affiliation{INAF/Brera Astronomical Observatory, via Bianchi 46, I-23807 Merate (LC), Italy}
\affiliation{Racah Institute of Physics, The Hebrew University, Jerusalem, 91904, Israel}

\author{Bin~Yang}
\affiliation{Key Laboratory of Dark Matter and Space Astronomy, Purple Mountain Observatory, Chinese Academy of Sciences, Nanjing 210008, China }
\affiliation{University of Chinese Academy of Sciences, Yuquan Road 19, Beijing, 100049, China}
\author{Zhi-Ping Jin}
%$^\ast$}
\affiliation{Key Laboratory of Dark Matter and Space Astronomy, Purple Mountain Observatory, Chinese Academy of Sciences, Nanjing 210008, China }
\author{Xiang Li}
\affiliation{Key Laboratory of Dark Matter and Space Astronomy, Purple Mountain Observatory, Chinese Academy of Sciences, Nanjing 210008, China }
\affiliation{University of Chinese Academy of Sciences, Yuquan Road 19, Beijing, 100049, China}
\author{Stefano Covino}
\affiliation{INAF/Brera Astronomical Observatory, via Bianchi 46, I-23807 Merate (LC), Italy}
\author{Xian-Zhong Zheng}
\affiliation{Key Laboratory of Dark Matter and Space Astronomy, Purple Mountain Observatory, Chinese Academy of Sciences, Nanjing 210008, China }
\author{Kenta Hotokezaka}
\affiliation{Racah Institute of Physics, The Hebrew University, Jerusalem, 91904, Israel}
\author{Yi-Zhong Fan}
%$^\ast$}
\affiliation{Key Laboratory of Dark Matter and Space Astronomy, Purple Mountain Observatory, Chinese Academy of Sciences, Nanjing 210008, China }
 \author{Tsvi Piran}
%$^\ast$}
\affiliation{Racah Institute of Physics, The Hebrew University, Jerusalem, 91904, Israel}
 \author{Da-Ming Wei}
\affiliation{Key Laboratory of Dark Matter and Space Astronomy, Purple Mountain Observatory, Chinese Academy of Sciences, Nanjing 210008, China }

\begin{abstract}
Long-duration ($>2$ s) $\gamma$-ray bursts that are believed to originate from the death of massive stars are expected to be accompanied by supernovae.  GRB 060614, that lasted 102 s, lacks a supernova-like emission down to very stringent limits and its physical origin is still debated. Here we report the discovery of near-infrared bump that is significantly above the regular decaying afterglow. This red bump is inconsistent with even the weakest known supernova.  However, it can arise from a Li-Paczy\'{n}ski macronova $-$ the radioactive decay of debris following a compact binary merger. If this interpretation is correct GRB 060614 arose from a compact binary merger rather than from the death of a massive star and it was a site of a significant production of heavy r-process elements. The significant ejected mass favors a black hole-neutron star merger but a double neutron star merger cannot be ruled out.
\end{abstract}
%\pacs{95.85.Ry, 95.85.Hp, 98.70.Sa}
\maketitle

%%%%%%%%%%%%%%%%%%%%%%%%%%%%%%%%%%%%%%%%%%%%%%%%%%%%%%%%%%%%%%%%%%%
Long-duration ($>2$ s) $\gamma$-ray bursts (GRBs) are believed to originate from Collapsars that involve death of massive stars and are expected to be accompanied by luminous supernovae (SNe). GRB 060614 was a nearby burst with a duration of 102 s at a redshift of 0.125 \cite{Gehrels2006}.
While it is classified as a long burst according to its duration,
{extensive searches did not find any} SNe-like emission
down to limits hundreds of times fainter than SN 1998bw \cite{Galama1998}, the archetypal hypernova that accompanied long GRBs \cite{Fynbo2006,Della2006,Gal-Yam2006}.
Moreover, the temporal lag and peak luminosity of GRB 060614 fell entirely within the short duration
subclass and the properties of the host galaxy distinguish it from other long-duration GRB hosts. Thus, GRB 060614 did not fit into the standard picture in which long duration GRBs arise from the collapse of massive stars while short ones arise from compact binary mergers.  It was nicknamed the ``long-short burst" as its origin was unclear.
Some speculated that it originated  from compact binary merger and thus it is  intrinsically a ``short" GRB \cite{Gehrels2006,Gal-Yam2006,Zhang2007,Barkov2009,Caito2011}.
Others proposed that it was formed in a new type of a Collapsar which produces an energetic $\gamma-$ray burst that is not accompanied by an SNe \cite{Fynbo2006,Della2006,Gal-Yam2006}.

Two recent developments may shed a new light on the origin of this object.
The first is the detection of a few very weak  SNe (e.g. SN 2008ha \cite{Valenti2009}) with  peak bolometric luminosities as low as $L \sim 10^{41}~{\rm erg~s^{-1}}$.
The second is the detection of an infrared bump, again with a $L \sim 10^{41}~{\rm erg~s^{-1}}$, in the late afterglow of the short burst GRB 130603B \cite{Tanvir2013,Berger2013}.
This  was interpreted  as a Li-Paczy\'{n}ski macronova (also called kilonova)
\cite{Li1998,Kulkarni2005,Rosswog2005,Metzger2010,Korobkin12,Barnes2013,Tanaka2013,Grossman2013}
$-$a near-infrared/optical transient
powered by the radioactive decay of heavy elements synthesized in the ejecta
 of a compact binary merger.
Motivated by these discoveries, we re-examined the afterglow data of this peculiar burst
searching for a signal characteristic to one of these events.

The  X-ray and UV/optical afterglow data of GRB 060614,  were extensively examined in the literature \cite{Mangano2007,Xu2009} and found to follow very well the fireball afterglow model up to $t\sim 20$ days \cite{Sari1998}. The $J$-band
has been disregarded because  only  upper limits $\sim 19-20^{\rm th}$ mag with a sizeable scatter are available at $t>2.7$ day, and these are too bright to significantly constrain even supernovae as luminous as SN 1998bw \cite{Cobb2006}. Here we focus on the optical emission. We  have re-analyzed all the late time (i.e., $t\geq 1.7$ day) Very Large Telescope (VLT) $V$, $R$ and $I-$band archival data and  the HST F606W and F814W archival data, including those reported in the literature \cite{Della2006,Gal-Yam2006} and several unpublished data points.
Details on data reduction are given in the Appendix.
Fig.\ref{fig:1} depicts the most complete late-time optical light curves (see Tab.\ref{Logtable} in the Appendix; the late VLT upper limits are not shown in Fig.\ref{fig:1}) of this burst.

The VLT $V$, $R$ and $I-$band fluxes decrease with time as $\propto t^{-2.30\pm 0.03}$ (see Fig.\ref{fig:1}, in which the VLT $V/I$ band data have been calibrated to the F606W/F814W filters of HST with proper $k-$corrections), consistent with that found earlier \cite{Della2006,Mangano2007,Xu2009}. However, the first HST F814W data point is significantly above the same extrapolated power-law decline. The significance of the deviation is  $\sim 6 \sigma$ (see the estimate in the Appendix).
No statistically-significant excess is present in both the F606W and the $R$ bands.
The F814W-band excess is made most forcibly by considering the color evolution of the transient, defined as the difference between the magnitudes in each filter, which evolves from $V-I \approx 0.65$ mag by the VLT (correspondingly for HST we have ${\rm F606W-F814W}\approx 0.55$ mag) at about $t\sim 1.7$ day to ${\rm F606W-F814W\approx 1.5}$ mag  by HST at about $13.6$ day after the trigger of the burst. With proper/minor extinction corrections, the optical to X-ray spectrum energy distribution for GRB 060614 at the epoch of
$\sim 1.9$ day is nicely fitted by a single power-law $F_\nu \propto \nu^{-0.8}$ \cite{Della2006,Mangano2007,Xu2009}. In the standard external forward shock afterglow model, the cooling frequency is expected to drop with time as $\nu_{\rm c}\propto t^{-1/2}$ \cite{Sari1998}. Thus, it cannot change the optical spectrum in the time interval of $1.9-13.6$ day. Hence, the remarkable color change and the F814W-band  excess of $\sim 1~{\rm mag}$ suggest a new component. Like in GRB 130603B this component was observed at one epoch only.  After the subtraction of the power-law decay  component, the flux of the excess component decreased  with time faster than $t^{-3.2}$ for $t>13.6$ days
An unexpected optical re-brightening was also detected in GRB080503, another `long-short' burst \cite{Perley2009}. However, unlike the excess component identified here, that re-brightening was achromatic in optical to X-ray bands and therefore likely originated by a different process.

%*****************************Fig.1***************************************
\begin{figure}
\includegraphics[width=140mm,angle=0]{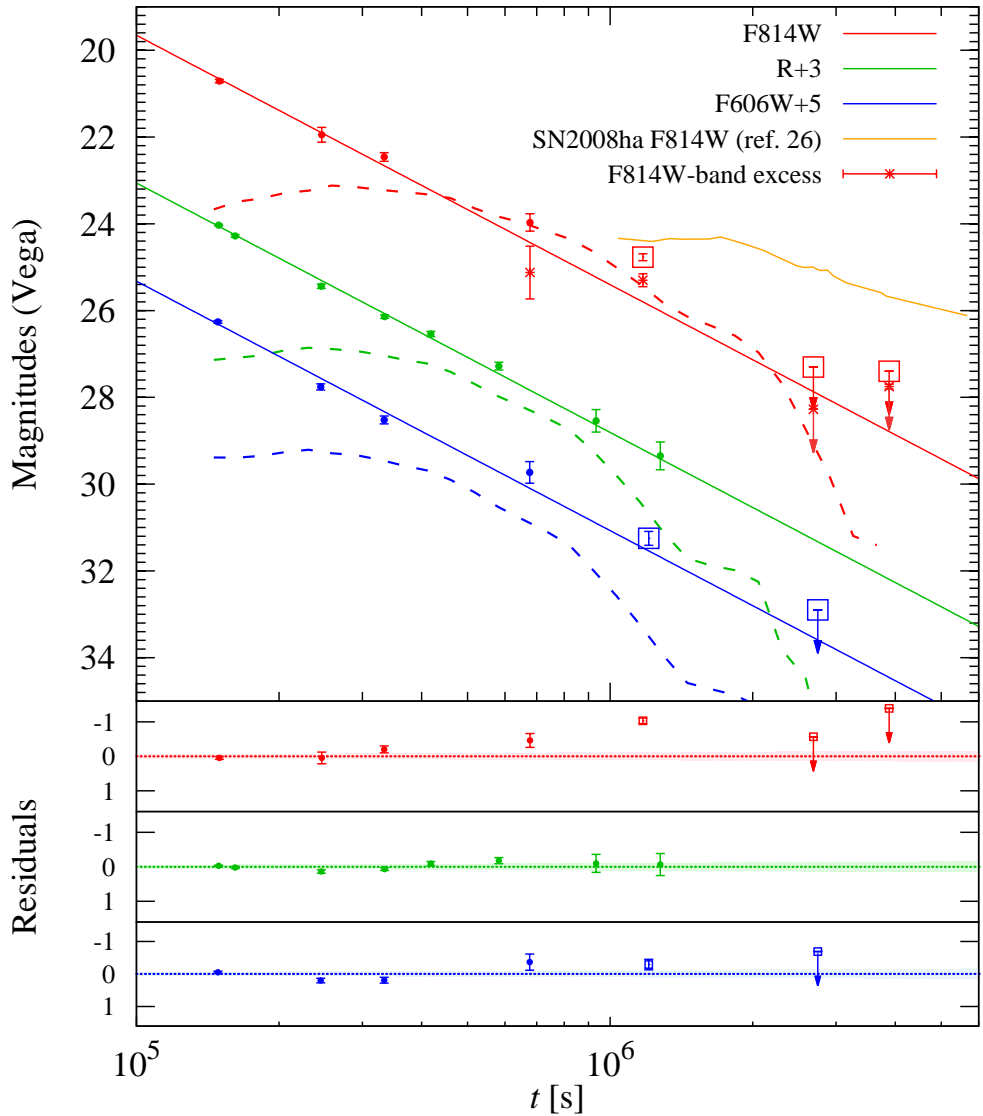}
\caption{The afterglow emission, not corrected for the small amount of foreground and  host extinction, of GRB 060614. Note that the VLT $V/I$ band data have been calibrated to the HST F606W/F814W filters with proper $k-$corrections (see the Appendix). The VLT data (the circles) are canonical fireball afterglow emission while the HST F814W detection (marked in the square) at $t\sim 13.6$ day is significantly in excess of the same extrapolated power-law decline (see the residual), which is at odds with the afterglow model.
%The dashed line represents the the ``excess component".
The F814W-band lightcurve of SN 2008ha expected at $z=0.125$ is also presented for comparison. The dashed lines are Macronova model light curves generated from numerical simulation \cite{Tanaka2014} for the ejecta from a black hole$-$neutron star merger.} \label{fig:1}
\end{figure}
%*************************************************************************

\begin{figure}
%\begin{picture}(0,600)
\includegraphics[width=140mm,angle=0]{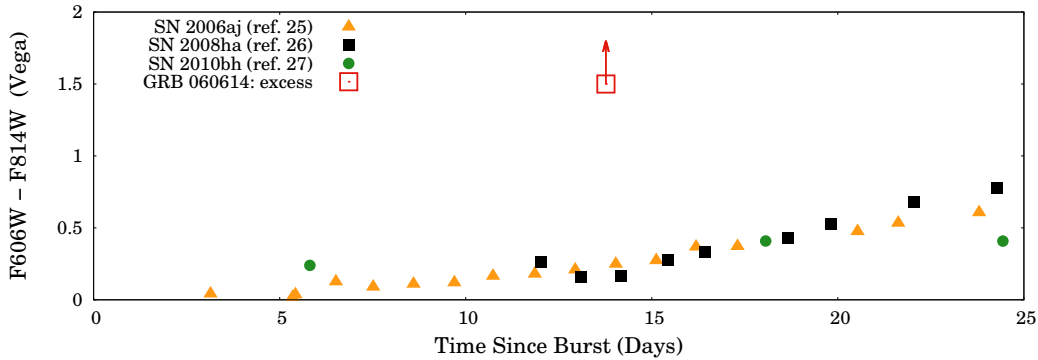}
%\put(0,0){\special{psfile=Figure-2Fan.eps angle=-0 hoffset=-240 voffset=-30
%hscale=80 vscale=80}}
%\end{picture}
\caption{The color (i.e., F606W-F814W) change of SN 2006aj, SN 2008ha, SN 2010bh and the ``excess component" identified in this work. The emission of these three supernovae, adopted from the literature \cite{Ferrero2006,Foley2009,Cano2011} has been shifted to $z=0.125$, the redshift of GRB 060614, with corrections on the time, frequency and extinction. Note that the ``excess component" is much redder than them.} \label{fig:2}
\end{figure}

Shortly after the discovery of GRB 060614 it  was speculated that it is powered by an ``unusual" core collapse of a massive star \cite{Fynbo2006,Della2006}. We turn now to explore whether the F814W-band excess can be powered by a weak supernova.
Fig.\ref{fig:2} depicts the color F606W$-$F814W of the excess component (we take F606W$-$F814W$\approx 1.5$ mag as a conservative lower limit of the color of the ``excess" component due to the lack of simultaneous excess in F606W-band) with that of SN 2006aj \cite{Ferrero2006}, SN 2010bh \cite{Cano2011} and SN 2008ha, an extremely dim event \cite{Foley2009}. The excess component has a much redder spectrum than the three supernovae. If the ``excess component" was thermal
%, as in the supernova radiation model,
it  had  a low effectively temperature $T_{\rm eff}< 3000~{\rm K}$ to yield the very soft spectrum.
Such unusually low effective temperature is also needed to account for the very rapid decline of the excess component. The expansion velocity can be estimated as $v \sim 1.2\times 10^{4}~{\rm km~s^{-1}}~(L/10^{41}~{\rm erg~s^{-1}})^{1/2}(T_{\rm eff}/3000~{\rm K})^{-2}(t/13.6~{\rm day})^{-1}$. The  implied $^{56}$Ni mass is $\sim 10^{-3}~M_\odot$ if this was a supernova-like event that peaked at $\sim 13.6$ day \cite{Valenti2009}. In this work we take a standard cosmology model with $H_0=71~{\rm km~s^{-1}~Mpc^{-1}}$, $\Omega_{\rm M}=0.32$ and $\Omega_\Lambda=0.68$.  Comparing with the extremely-faint SN 2008ha after proper corrections to $z=0.125$, the peak F814W-band emission of the ``excess component" is lower by $\sim 1$ mag and the decline is also much faster. Hence the ``excess component" is remarkably different from SN 2008ha.

%*****************************Fig.2***************************************
%*************************************************************************

The low luminosity as well as the  low effective temperature of the transient emission are typical characteristics of a Macronova, a transient arising from the radioactive $\beta$-decay of material ejected in
 a compact binary merger. The opacity of the macronova material is determined by the Lanthanides that are   produced via  $r-$process
 in the neutron-rich outflow. This opacity is very large ($\kappa \approx 10~{\rm cm^{2}~g^{-1}}$) resulting in a weak, late  and red emission.  The emerging flux is greatly diminished by line blanketing, with the radiation peaking in the near-infrared and being produced over a timescale of $\sim 1-2$ weeks \cite{Barnes2013,Tanaka2013}.
Simple analytic estimates, using a radioactive $\beta$-decay heating rate of  $10^{10}~{\rm erg~s^{-1}~g^{-1}}~[t/(1+z){\rm 1~day}]^{-1.3}$ \cite{Korobkin12,Tanaka2014},
suggest that in order to explain the observed F814W-band excess,
the required ejecta mass and expansion velocity are:
$M_{\rm ej}\sim 0.13~M_{\odot}~(L/10^{41}~{\rm erg~s^{-1}})(t/13.6~{\rm day})^{1.3}$ and
$v\sim 0.1c~(L/10^{41}~{\rm erg~s^{-1}})^{1/2}(T_{\rm eff}/2000~{\rm K})^{-2}(t/13.6~{\rm day})^{-1}$, respectively. Note that the Macronova outflow is quite cold at such a late time \cite{Barnes2013,Tanaka2013} .
The effective temperature is  $T_{\rm eff}  \approx 2000$ K and the observer's F814W-band is above the peak of the black body spectrum. The emitting radius and the corresponding expansion velocity are much larger than in a supernova at this stage.
Scaled up numerical simulations of lighter ejecta from black hole$-$neutron star mergers \cite{Tanaka2014} suggest that $M_{\rm ej}\sim 0.1~M_\odot$ and a velocity $\sim 0.2c$ can account for the observed F814W-band excess. This numerical example is presented in Fig.\ref{fig:1} in dashed lines.

The implied ejecta mass is large compared with the mass ejection estimated numerically to take place in double neutron star mergers. However, it is within the possible range of dynamical ejecta of black hole-neutron star mergers with some extreme parameters (a large neutron star radius and a high black-hole spin aligned with the orbital angular momentum) \cite{Rosswog2005,Piran+13,lovelace2013,Foucart2014,kyutoku2015}.  An accretion disk wind may contribute some additional mass as well \cite{Metzger2010,Just2015,Fernandez2014}.
However, the radioactive heating due to fission of the heavy $r-$process nuclei, which is quite uncertain and
subdominant in current heating estimates \cite{Korobkin12}, may play an important role in the energy deposition.  It
may increase the energy deposition rate at around ten days by a significant factor \cite{Wanajo2014}.
This may reduces the required ejecta mass to $\sim 0.03-0.05~M_\odot$. This range of the ejecta masses is well within the range of the dynamical ejecta of black-hole neutron star mergers and it is even compatible with some estimates of double neutron star mergers.

We conclude that while a weak supernova cannot explain the observations, a high mass ejection macronova
may. Like in GRB 130603B we must caution here that this interpretation is based on a single data point. However, if this interpretation is correct, it has far reaching implications. First, the presence of macronovae
in both the canonical short burst GRB 130603B and in this  ``long-short" one,
GRB 060614, suggests that the phenomenon is common and the prospects of detecting these transients are promising.
A more conclusive detection based on more than a single data point could be achieved in the future provided that denser HST observations are carried out. Moreover, as a black hole--neutron star merger is favored in explaining the large ejected mass this implies that such binary systems may exist and their mergers are also responsible for GRBs. It also suggests that the ``long-short" burst was in fact ``short" in nature, namely, it arose from a merger and not from a Collapsar. The fact that  a merger
generates a 100 sec long burst is interesting and puzzling by itself.

%The ejection of so much mass makes these
%compact-object mergers sites of significant production of $r-$process elements.
Clearly such events would contribute a significant fraction of the $r-$process material\ref{Lattimer1974}. The actual contribution
relative to the contribution of 130603B-like events is difficult to estimate as it is unclear which fraction of the
Macronovae/kilonovae behave as each type. Because of beaming most
mergers will not be observed as GRBs. However, they emit omnidirectional gravitational radiation that can be detected by the upcoming Advanced LIGO/VIRGO/KAGRA detectors. These near-infrared/optical macronovae could serve as promising electromagnetic counterparts of gravitational wave triggers in the upcoming Advanced LIGO/VIRGO/KAGRA era.\\

{\it Acknowledgments.}
We thank Dr. A. Gal-Yam for communication and the referees for helpful comments. This work was supported in part by the National Basic Research Programme of China (No. 2013CB837000 and No. 2014CB845800), NSFC under grants 11361140349, 11103084, 11273063, 11303098 and 11433009, the Foundation for Distinguished Young Scholars of Jiangsu Province, China
(Grant No. BK2012047),  the Chinese Academy of Sciences via the 100 talent program and the Strategic Priority Research Program (Grant No. XDB09000000), the Israel ISF$-$China BSF grant and the I-Core center for excellence ``Origins" of the ISF. SC has been supported by ASI grant I/004/11/0.

{\it Author contribution.} PMO (Z.P.J, B.Y, X.L and X.Z.Z) and INAF/OAB (S.C) carried out the data analysis, following Y.Z.F's suggestion. HU (K.H and T.P) and PMO (Y.Z.F, D.M.W and Z.P.J) led the interpretation of the data. Y.Z.F and T.P prepared the paper and all authors joined the discussion.

{\it Author information.} Correspondence should be addressed to Y. Z.~Fan (e-mail: yzfan@pmo.ac.cn), Z. P. Jin (e-mail: jin@pmo.ac.cn) and T. Piran (e-mail: tsvi.piran@mail.huji.ac.il).\\

%$^\ast$Corresponding authors.\\
%Electric addresses: yzfan@pmo.ac.cn (YZF), jin@pmo.ac.cn (ZPJ), tsvi.piran@mail.huji.ac.il (TP).

%\headertitle{Methods}
%\mainauthor{Yang et al.}

\clearpage
\begin{center}
{\Large Appendix} \\
\end{center}

%\headertitle{Appendix}
%\mainauthor{Yang et al.}

\section{Data reduction}

We retrieved the public VLT imaging data of GRB\,060614 from ESO Science Archive Facility (http://archive.eso.org).
The raw data were reduced following standard procedures, including bias subtraction, flat fielding, bad pixel removal, and combination.
Observations made with the same instrument and filter at different epochs are compared to that of the last epoch. The software package ISIS (http://www2.iap.fr/users/alard/package.html) is used to subtract images and measure the GRB afterglow from the residual images.
Photometric errors are estimated from the photon noise and the sky variance to $1\sigma$ confidence level.
The 3\,$\sigma$ of the background RMS of the residual images is taken as the limiting magnitude.
Finally, standard stars observed on Jun 16, 2006 were used for the absolute calibration. The results are shown in Tab.\ref{Logtable}.
%These VLT data are well following a single power-law decay in all bands.
We assumed that the afterglow is characterized by the same power-law spectrum with index $\beta=0.80$ \cite{Mangano2007} during these observations, with which we get the $k$-corrections between the VLT $V/I$ and HST F606W/F814W magnitudes, namely 0.12 mag and 0.02 mag, respectively. Such corrections had been taken into account in Fig.\ref{fig:1}.

\begin{figure}
\includegraphics[width=140mm,angle=0]{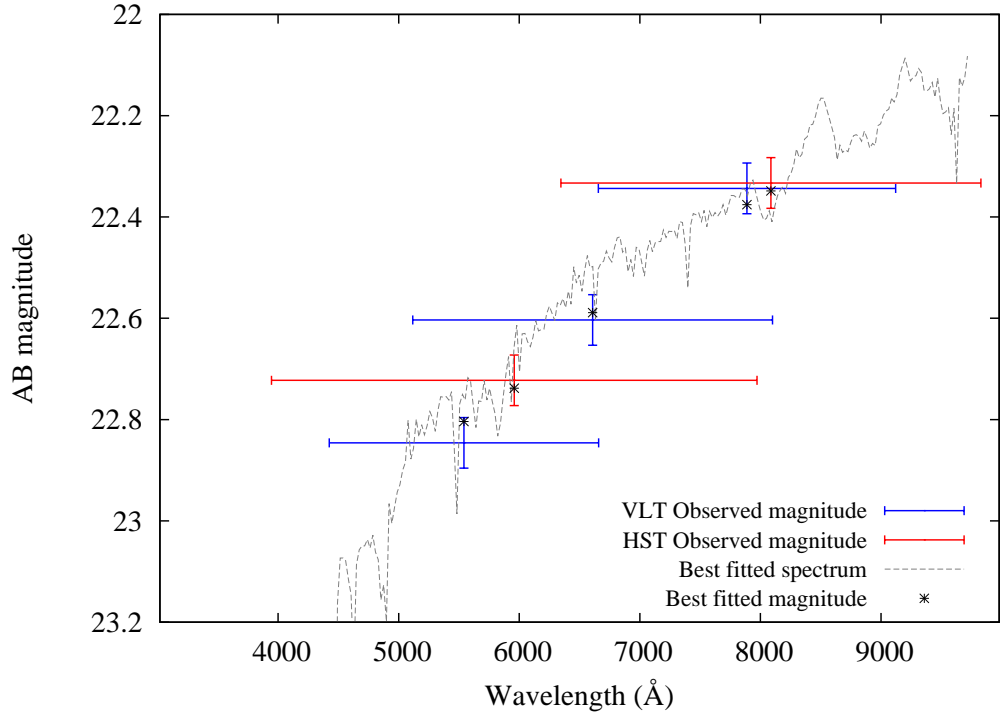}
\caption{Following Gal-Yam et al. (2006)\cite{Gal-Yam2006}, we fit the host galaxy magnitudes by an Sc type template. The redshift of the host galaxy $z=0.125$ and the Galactic extinction $A_{\rm V}=0.07$ mag have been taken into account.} \label{figurehost}
\end{figure}

HST archive data of GRB\,060614 are available from the Mikulski Archive for Space Telescopes (MAST; http://archive.stsci.edu),
including one observation with WFPC2 and four observations with ACS in F606W and F814W bands. The reduced data provided by MAST were used in our analysis. The last visit is taken as the reference and the other images of the same filter are subtracted in order to directly measure fluxes of the afterglow from the residual images. Empirical PSFs was built with bright stars in each image. Bright compact objects in the same field were used to align and relatively calibrate these images.  WFPC2 image differs from ACS image in PSF. Before image subtraction, the WFPC2 and ACS images were matched to the same resolution by convolving each with the other's PSF. The PSF-matched WFPC2 and ACS images were aligned and subtracted. Aperture photometry was carried out for the afterglow in the residual image. The aperture correction derived from the empirical PSF was applied to yield the total flux.
The host galaxy was used to relatively calibrate
the afterglow between images, and the ACS zeropoints were used for absolute calibration.
%The host galaxy was used to relatively calibrate the afterglow between images,
%and the ACS zeropoints of the reference images were used for a detection.
%to calculate the magnitude of the afterglow.
If the signal of the afterglow is too faint to be a secure detection, an upper limit of 3\,$\sigma$ background RMS is adopted. The magnitudes of the host galaxy are measured in the last observation of all filters and can well be fitted by an Sc type galaxy template, see Fig.\ref{figurehost}, demonstrating the self-consistence of our results.
Our results are summarized in Tab.\ref{Logtable}, being well consistent with these given by other groups \cite{Della2006,Gal-Yam2006,Xu2009}.\\

\begin{table}
\caption{Log of observations}.
\label{Logtable}
\begin{tabular}{lllll}
\hline \hline
Time from GRB$^{a}$    & Filter & Exposure time     & Instrument & Vega Magnitude$^{b}$\\
(days)    &	& (s)		&	  &		 \\ \hline \hline
1.72034   & $V$	& 2$\times$120  & VLT+FORS1  & 21.38$\pm$0.03\\
2.83515   & $V$	& 4$\times$90	& VLT+FORS1  & 22.88$\pm$0.07\\
3.86077   & $V$	& 2$\times$120+4$\times$180  & VLT+FORS1  & 23.64$\pm$0.09\\
7.82790   & $V$	& 3$\times$180  & VLT+FORS2  & 24.85$\pm$0.25\\
23.79964   & $V$	& 2$\times$120  & VLT+FORS1  & $>$24.8\\
32.78644   & $V$	& 5$\times$120  & VLT+FORS1  & [$22.96\pm0.04$] \\
108.57188   & $V$	& 3$\times$300  & VLT+FORS2  & [$22.87\pm0.04$] \\  \hline
1.72583   & $R$	& 2$\times$120  & VLT+FORS1  & 21.03$\pm$0.02\\
1.86974   & $R$	& 2$\times$120  & VLT+FORS1  & 21.28$\pm$0.03\\
2.84199   & $R$	& 2$\times$120  & VLT+FORS1  & 22.44$\pm$0.05\\
3.86899   & $R$	& 2$\times$120+4$\times$180  & VLT+FORS1  & 23.14$\pm$0.04\\
4.84365   & $R$	& 2$\times$180  & VLT+FORS1  & 23.54$\pm$0.06\\
6.74083   & $R$	& 3$\times$180  & VLT+FORS1  & 24.28$\pm$0.09\\
10.81441   & $R$	& 2$\times$300  & VLT+FORS1  & 25.54$\pm$0.26\\
14.77259   & $R$	& 4$\times$300+4$\times$180 & VLT+FORS1  & 26.35$\pm$0.32\\
19.67818   & $R$	& 6$\times$240  & VLT+FORS1  & $>$26.3 \\
23.80494   & $R$	& 2$\times$120  & VLT+FORS1  & $>$24.6 \\
32.79667   & $R$	& 3$\times$180  & VLT+FORS1  & $>$25.4 \\
44.73601   & $R$	& 5$\times$240+2$\times$180  & VLT+FORS1  & $>$26.3 \\
64.70367   & $R$	& 12$\times$300 & VLT+FORS1  & [$22.42\pm0.03$]\\ \hline
1.73236   & $I$	& 3$\times$120  & VLT+FORS1  & 20.73$\pm$0.04\\
2.84826   & $I$	& 3$\times$120  & VLT+FORS1  & 21.97$\pm$0.17\\
3.85840   & $I$	& 4$\times$300  & VLT+FORS1  & 22.48$\pm$0.10\\
7.84052   & $I$	& 3$\times$120  & VLT+FORS2  & 23.99$\pm$0.20\\
23.81008   & $I$	& 2$\times$120  & VLT+FORS1  & $>$23.9\\
32.80572   & $I$	& 3$\times$180  & VLT+FORS1  & [$21.99\pm0.04$] \\
108.58482   & $I$	& 4$\times$240  & VLT+FORS2  & [$21.94\pm0.04$] \\ \hline
13.97023	  & F606W & 	 6000	    & HST+WFPC2  & 26.25$\pm$0.16 \\
31.76674   & F606W	&    3600       & HST+ACS    & $>$27.9   \\
85.59018   & F606W	&    4372       & HST+ACS    & [$22.66\pm0.02$]  \\ \hline
13.57128   & F814W& 	 6000    & HST+WFPC2  & 24.77$\pm$0.08  \\
31.09855   & F814W&    3600       & HST+ACS    & $>$27.3   \\
44.95641   & F814W&    4840       & HST+ACS    & $>$27.4   \\
139.44208  & F814W&    4840       & HST+ACS    & [$21.95\pm0.02$]  \\ \hline
\end{tabular}

\textbf{Notes.}

a. Time since the burst trigger, which occurred at 2006 June 14, 12:43:48 UT.

b. These values have not been corrected for the Galactic extinction of $A_{\rm V}=0.07$ mag. The magnitudes in square brackets are for the host galaxy. The definitions of the errors and upper limits are described in the text (note that 0.02 magnitude of uncertainty in absolutely calibrations has been added to the statistical errors).
\end{table}

%*****************************Fig.3***************************************
%*************************************************************************

\section{The decline rate of the VLT afterglow data and the significance of the excess}

As found in previous studies, the late time optical/X-ray afterglow emission of GRB 060614 can be interpreted within the fireball forward shock model \cite{Mangano2007,Xu2009}. Motivated by such a fact, we assume that the I, R, V lightcurves follow the same power-law decline. In our fit there are four free parameters, three are related to the initial flux/magnitude in these three bands and the last is the decline rate needed in further analysis. We fitted all the VLT data (combined I, R, V band together) during the first 15 days (after which there are just upper limits) to determine these four parameters as well as their errors. The best-fit decline is found to be $\propto t^{-2.3\pm 0.03}$, well consistent with that obtained in optical to X-ray bands in previous studies \cite{Della2006,Mangano2007,Xu2009}. As a result of the propagation of uncertainties, the errors of the best-fit light curves are consequently inferred (the shadow regions in the residual plot of Fig.1 represent the $1\sigma$ errors of the best-fit light curves). Please note that in Fig.1 the VLT V/I band emission have been calibrated to HST F606W/F814W filters with proper $k-$corrections. The flux separation between the HST F814W-band data and the fitted curve at $t\sim 13.6$ day is $F_{\rm excess}=0.182~{\rm \mu Jy}$. The flux error of the F814W-band emission at $t\sim 13.6$ day is $\delta F_{\rm obs}\approx 0.024~{\rm \mu Jy}$. The flux error of the best fitted F814W-band lightcurve at $t\sim 13.6$ days is $\delta F_{\rm fit}\approx 0.012~{\rm \mu Jy}$. The significance of the excess component is estimated by ${\cal R}= F_{\rm excess}/\sqrt{\delta F_{\rm obs}^{2}+\delta F_{\rm fit}^{2}} \sim 6$. We therefore suggest that the excess component identified in this work is statistically significant at a confidence level of $\sim 6 \sigma$.

\end{document}